# Mechanisms of GaN quantum dot formation during nitridation of Ga droplets

H. Lu[1], C. Reese[1], S. Jeon[1], A. Sundar[1], Y. Fan[2], E. Rizzi[1], Y. Zhuo[2], L. Qi[1] and R. S. Goldman[1,3*]

*1Department of Materials Science and Engineering
University of Michigan, Ann Arbor, MI 48109-2136
2Key Laboratory for Thermal Science and Power Engineering of the Ministry of Education,
Dept. of Energy and Power Engineering, Tsinghua University., Beijing 100084
3Department of Physics
University of Michigan, Ann Arbor, MI 48109-2136*

(11/4/19)

## Abstract

We have examined the formation mechanisms of GaN quantum dots (QDs) via annealing of Ga droplets in a nitrogen flux. We consider the temperature and substrate dependence of the size distributions of droplets and QDs, as well as the relative roles of Ga/N diffusivity and GaN nucleation rates on QD formation. We report on two competing mechanisms mediated by Ga surface diffusion, namely QD formation at or away from pre-existing Ga droplets. We discuss the relative roles of nucleation and coarsening dominant growth, as well as the polytype selection, on various substrates. The new insights provide an opportunity for tailoring QD size and polytype distributions for a wide range of III-N semiconductor QDs.

---

1) H. Lu and C. Reese contributed equally to this work.

2) Corresponding Author: rsgold@umich.edu

In recent years, quantum dots (QDs) based on gallium nitride (GaN) and its alloys have been demonstrated for a wide variety of device applications, such as solar cells,[1,2] light-emitting diodes,[3,4,5] lasers,[6,7] and single-photon emitters.[8,9] QDs are typically grown epitaxially via a strain-induced Stranski-Krastanov (S-K) growth mode transition, which leads to a misfit-strain induced polarization in the QDs.[10] On the other hand, the nucleation and conversion of QDs via nitridation of metallic droplets, known as droplet epitaxy (DE), has attracted much attention as misfit-strain-induced-polarization is expected to be minimized. To date, DE of GaN QDs has been demonstrated on a variety of substrates, including 6H-SiC(0001),[11] Si(111),[12,13,14] AlGaN/6H-SiC(0001),[15] and c-$Al_2O_3$,[16] with single electron transistor achieved on AlN/3C-SiC(001).[17] Furthermore, understanding of DE is critical for elucidation of the mechanisms for GaN growth under Ga-rich conditions, which are often argued to be a version of liquid phase epitaxy.[18] Indeed, during Ga-rich GaN growth, the low solubility of carbon in Ga and the reactivity of oxygen in Ga, with subsequent desorption of GaO at growth temperatures, leads to low carbon[19] and oxygen[20] co-incorporation. Therefore, DE GaN QDs are expected to be superior to those grown by the SK method.[21]

However, conflicting results have been reported regarding the formation mechanisms of DE GaN QDs. For example, Wang *et al.*[17] and Gherisimova *et al.*[22] reported on the formation of QDs via a liquid phase epitaxy (LPE)-like process, where GaN crystallizes along the substrate/droplet interface when N supersaturates the liquid Ga. On the other hand, Debnath *et al.*[15] proposed a surface diffusion-driven mechanism, where Ga diffuses away from the droplets and reacts with N on the surface to form small QDs. Finally, Kawamura *et al.*[23] and Otsubo *et al.*[24] propose a formation mechanism where N diffuses along the surface to the droplet edges, and small QDs nucleate at the periphery. Here, we investigate the formation mechanisms for DE GaN QDs using a combined computational-experimental approach. Our first-principles calculations of activation barriers suggest that N is immobile while Ga has a relatively high surface diffusivity, independent of the starting surface structure and chemistry. We present the temperature and substrate dependence of the droplet and QD size distributions and report on two competing mechanisms mediated by Ga surface diffusion, namely QD formation at or away from pre-existing Ga droplets. We also report the formation of zincblende vs. wurtzite polytype GaN and discuss the relative roles of nucleation and coarsening dominant growth, as well as the polytype selection, on various substrates. These mechanisms provide an opportunity for tailoring QD size and polytype distributions for a wide range of III-N semiconductor QDs.

The QD arrays were prepared using molecular beam epitaxy (MBE), with solid Ga and RF-plasma assisted nitrogen sources. Si(001) and Si(111) substrates were etched in a 5% HF solution for 1 minute, followed by insertion into the load-lock chamber within 30 minutes. Following UHV transfer into the MBE, the Si substrates were high temperature annealed (substrate temperature of 900 °C) for 10 minutes to desorb native oxides. For select Si(001) substrates, the high temperature annealing step was omitted, in order to achieve a native oxide surface. Next, with the substrate temperature set to 550 °C, the substrate was exposed to a N flux of $1.0 \times 10^{-6}$ Torr for 10 minutes, the "initial" nitridation step, which provides surface $SiN_x$. To form droplets, an equivalent Ga thickness of 7.5 ML (~1 nm) was deposited at a rate of 0.75 ML/s. For the "final" nitridation, the N shutter was opened, while the Ga shutter was simultaneously closed. Meanwhile, the temperature was either held constant at 550 °C ("fixed") or increased at 50°C/min to 650°C ("moderate") or 720°C ("high"), with the duration of temperature ramping less than 4 minutes. During the "final" nitridation, the sample was exposed to a N flux of $1.0 \times 10^{-6}$ Torr for 30 minutes. For all growths, the RF-plasma source power and $N_2$ flow rate were fixed at 350 W and 1.0 sccm, respectively, yielding a N flux of $1.0 \times 10^{-6}$ Torr [as determined by the partial pressure of 14 amu with a residual gas analyzer (RGA)]. The 1 sccm flow rate corresponds to the maximum flux available for our N source configuration. The surface morphologies of the Ga droplet and GaN QD arrays were examined *ex-situ* using atomic force microscopy (AFM) in non-contact mode with etched Si tips.

The diffusion barriers for Ga and N on silica and silicon surfaces were determined using first-principles calculations with density functional theory (DFT) in the generalized-gradient approximation (GGA), as discussed in the supplemental . In all cases, the activation energy for diffusion (or diffusion barrier) of N is predicted to be significantly higher than that for Ga. On silica surfaces, the computed diffusion barriers are 4.3 eV for N and 0.32 eV for Ga. Similarly, for Si (001) surfaces, the computed diffusion barriers are 2.64 eV for N and 0.38 eV for Ga. Finally, on Si(111) surfaces, the computed diffusion barriers are 3.44 eV for N and 0.41 eV for Ga. Thus Ga is expected to rapidly diffuse on silica and silicon surfaces, while N atoms are nearly immobile on all of those surfaces.

To determine the surface reconstructions prior to and during growth, reflection high-energy electron diffraction (RHEED) patterns were collected along the [110] direction, as shown in Fig. 1. For silica surfaces [Fig. 1(a)], as the temperature is increased to 550 °C, a streaky 1×1 pattern is apparent, suggesting an unreconstructed Si(001) surface with the presence of an oxide layer, which we term the "silica" surface. For the Si(001) surface [Fig. 1(b)], a 2×1 reconstruction appears as the temperature is increased to 900 °C, indicating an oxide-free Si(001) surface. For the Si(111) surface [Fig. 1(c)], as the temperature is increased to 900 °C, the appearance of a 7×7 reconstruction suggests an oxide-free Si(111) surface.

During the initial nitridation of silica, streaky RHEED patterns [Fig. 1(d)] resemble the pattern observed after surface preparation [Fig. 1(a)], suggesting incomplete nitrogen surface coverage. During the deposition of Ga, hazy-streaky RHEED patterns, with the underlying streaks corresponding to the 1×1 Si(001) [Fig. 1(g)], suggest the formation of Ga droplets in lieu of complete Ga surface coverage. During the final nitridation of the silica surface, shown in Fig. 1(g), the RHEED pattern transitions to concentric rings, containing both (WZ) and zincblende (ZB) GaN reflections, suggesting the conversion of Ga droplets into crystalline GaN with multiple polytype and/or orientations, presumably due to incomplete N coverage of the $SiO_2$ layer at the surface as shown in Fig. 1(h).

On silicon surfaces, hazy-streaky RHEED patterns [Fig. 1(e) and (f)] were observed after "initial" nitridation, suggesting that the surface is covered with an amorphous layer of $Si_xN_y$. During the deposition of Ga, hazy-streaky RHEED patterns, with the underlying streaks corresponding to the 1×1 Si(001) [Fig. 1(e)] or 7×7 Si(111) [Fig. 1(f)] surfaces are apparent, suggesting the formation of Ga droplet ensembles in lieu of complete Ga surface coverage. During the final nitridation on the Si(001) surface, shown in Fig 1(k), a cubic spotty pattern is observed with reflections corresponding to ZB GaN, suggesting the transformation of Ga droplets to ZB GaN QDs. Presumably, the formation of ZB QDs is due to epitaxial QD growth mediated by the cubic substrate surface.[12,15,17] During the final nitridation of the Si(111) surface, shown in Fig. 1(l), spotty-ring RHEED patterns corresponding to WZ GaN are observed, indicating the conversion of droplets to WZ GaN QDs. Presumably, the growth of WZ QDs arises due to epitaxial QD growth on the hexagonal-like substrate surface.[12,15,17]

For each surface, post-growth AFM images of Ga droplet ensembles and GaN QD ensembles nitridated at fixed, moderate, and high temperatures are presented in Figs. 2(a)-(d) [silica], 3(a)-(d) [Si(001)], 4(a)-(d) [Si(111)]. In addition, the corresponding Ga droplet and GaN QD size distributions are fit with Gaussian and/or log-normal functions, in order to extract the most probable droplet and QD diameters, $d_m$, as shown in Fig. 2(e), Fig. 3(e), and Fig. 4(e).

For all the Ga droplet ensembles, the size distributions are best described (i.e. $R^2 > 0.99$) with a lognormal function, consistent with the expected absence of coarsening during liquid-like droplet formation.[25] For Ga droplet ensembles on silica, Si(001), and Si(111) surfaces, $d_m$ values are 27±3 nm, 47±8 nm, and 47±4 nm with droplet densities of $1.6×10^{10}$ cm$^{-2}$, $3.7×10^9$

cm$^{-2}$, and $5.3 \times 10^9$ cm$^{-2}$, respectively. The lower values of droplet d$_m$ and density on the silica surface in comparison to those of the silicon surfaces is likely due to the "patchy" vs. complete coverage of Si$_x$N$_y$, as shown in Figs 5(a)-(b) and Fig 5(e)-(f) respectively. Presumably, the limited lateral extent of the Si$_x$N$_y$ patches on the silica surface inhibits Ga droplets nucleation and coalescence, thereby limiting the size of Ga droplets.

For most surfaces, following final nitridation, the QD d$_m$ values decrease while the QD densities increase in comparison to those of the Ga droplets. As illustrated in Fig 5(c), for those cases, the dominant QD formation mechanism is Ga out-diffusion, expected to be well-described by a log-normal distribution. Indeed, for most cases, fits to a log-normal distribution lead to R$^2$>0.99. On silicon surfaces, QD nucleation occurs anywhere along the surfaces SiN$_x$, as shown in Fig 5(g).

For silica surfaces, due to the limited Ga surface diffusion length, $\boldsymbol{\lambda_{Ga}}$, out-diffusing Ga atoms cannot reach nearby Ga droplets; instead, they nucleate at SiN$_x$ patches between Ga droplets, as shown in Fig 5(c). On the other hand, following final nitridation of the silica surface at high temperature, the QD d$_m$ values increase while the QD densities decrease in comparison to those of the Ga droplets. As illustrated in Fig 5(d), in this case, the QD formation mechanism includes coarsening well-described by either a Gaussian or Lorentzian distribution. Indeed, fits to a Gaussian distribution lead to R$^2$> 0.99 but those of a Lorentzian distribution leads to R$^2 \approx$ 0.96. Due to the longer $\boldsymbol{\lambda_{Ga}}$ at high final nitridation temperature, out-diffusing Ga atoms are able to reach other Ga droplets, resulting in droplet coarsening, as shown in Fig 5(d).

To understand the enhanced Ga surface diffusion on silica surfaces, we consider DFT computed Ga adsorption energies on nitrided and oxidized Si surfaces, as described in the Supplemental Materials. Due to the negative values of adsorption energies for Ga on nitrided Si surfaces, surface N is likely to inhibit the diffusion of Ga atoms. On the other hand, for silica surfaces, the initial nitridation leads to incomplete nitrogen surface coverage. Instead, due to the positive values of adsorption energies for Ga on oxidized silicon surfaces, the regions with oxygen may serve as fast diffusion path for Ga adatom, thereby facilitating coarsening.

For both Si(001) and Si(111), bimodal distributions of QD $d_m$ values are observed at moderate and high final nitridation temperatures. In addition, several of the QDs formed at the highest nitridation temperature exhibit a ring-shaped cluster morphology, suggesting that N incorporation, followed by QD nucleation, occurs at the periphery of the largest observed Ga droplets.[23] Indeed, the ring diameters are larger on Si(111) than on Si (001). Since the gallium out-diffusion which leads to ring formation occurs on $Si_xN_y$ surfaces, the higher Ga diffusion barrier on Si(111) vs. Si(001) may not have an impact on the ring formation. Similar morphologies have been reported in other III-V material systems, such as GaAs/GaAs[28,29] and InGaAs/GaAs,[30] where ring formation is attributed to simultaneous out-diffusion of Ga from the droplet center and arsenic crystallization at the edges. The observation of the ring-morphology with smaller individual QDs, as well as the appearance of a multi-modal QD distribution, as discussed above, suggests that Ga adatoms diffusing outwardly from the droplet center are able to nucleate as smaller QDs elsewhere on the surface.

Finally, on Si(111) surfaces, for the fixed final nitridation temperature, it is interesting to note that QDs are preferentially located at the edges of ~2 nm height macrosteps, often termed "mounds", consistent with nucleation of both $Si_3N_4$ and QDs at regions of positive curvature.[31,32]

In summary, we have examined the formation mechanisms of GaN quantum dot formation during nitridation of Ga droplets. We consider the temperature and substrate dependence of the size distributions of droplets and QDs, as well as the relative roles of Ga/N diffusivity and GaN nucleation rates on QD formation. We report on the polytype selection of ZB GaN QDs on Si(001), WZ GaN QDs on Si(111), and mixed ZB/WZ GaN QDs on silica surfaces. We also report on two competing mechanisms mediated by Ga surface diffusion, namely QD formation at or away from pre-existing Ga droplets. On silica surfaces, coarsening-dominant growth leads to coalescence of Ga droplets and consequently larger hemispherical shaped QD morphologies at high nitridation temperature. At low temperature, due to limited Ga diffusion length, QDs form in between Ga droplets. On Si(001) and Si(111) surfaces, nucleation-dominant growth leads to smaller QDs as GaN QDs can nucleate anywhere on the surface. These new insights provide an opportunity for tailoring QD size and polytype distributions for a wide range of III-N semiconductor QDs.

**Acknowledgement**

This work was supported by the National Science Foundation (NSF) through the Materials Research Science and Engineering Center (MRSEC) at the University of Michigan, under Grant No. DMR-1120923. H.L and E.R. were supported in part by NSF Grants ECCS-1610362 and DMR-1410282. Y.F. was supported in part by the China National Natural Science Foundation via project No. 51776107. CASTEP, the DFT module of Materials Studio, was provided by Prof. Liangliang Li at School of Materials Science and Engineering, Tsinghua University, Beijing, China. We thank Professor Theodore D. Moustakas for his very helpful discussions and suggestions.

**Figure Captions**

Fig. 1: RHEED patterns collected along the [110] axis for silica [(a),(d),(g),(j)], Si(001) [(b),(e),(h),(k)], and Si(111) [(c),(f),(i),(l)] surfaces. In the first row, following surface preparation, streaky patterns corresponding to diffraction from the (a) silica, (b) Si(001), and (c) Si(111) surfaces are apparent. In the second row, during surface nitridation, streaky patterns on (d) silica indicate that the surface is not nitridated due to the coverage of oxide. Diffused patterns on (e) Si(001) and (f) Si(111) indicate that an amorphous layer of $Si_xN_y$ is formed. In the third row, during Ga droplet formation, the RHEED patterns on (g) silica, (h) Si(001), and (i) Si(111), turn hazy due to partial Ga coverage, but still exhibit the underlying Si streaks. In the fourth row during nitridation of Ga droplets, the RHEED patterns reveal transitions to (j) polycrystalline GaN on silica, (k) ZB GaN on Si(001), and (l) WZ GaN on Si(111).

Fig. 2: AFM images of Ga droplet and GaN QD ensembles grown on silica: (a) Ga droplets deposited at 550 °C, (b) GaN QDs nitrided at 550 °C, (c) GaN QDs nitrided at 650 °C, and (d) GaN QDs nitrided at 720 °C. The color-scale ranges displayed are (a) 24.8 nm, (b) 17.6 nm, (c) 19.3 nm, and (d) 18.2 nm. The corresponding size distributions from images (a)-(d) are shown in (e) and the frequency is the percentage of QDs with diameters within a specified range. For Ga droplets, the $d_m$ value is 27±3 nm with a density of $1.6×10^{10}$ cm$^{-2}$. For GaN QDs nitrided at 550, 650, and 720 °C, the $d_m$ values (densities) are 18±3 nm ($8.3×10^{10}$ cm$^{-2}$), 21±2 nm ($4.0×10^{10}$ cm$^{-2}$), and 34±4 nm ($5.4×10^{9}$ cm$^{-2}$) respectively. (a)-(c) are fitted with a log-normal distribution where $R^2 > 0.99$, and (d) is fitted using a Gaussian distribution where $R^2 > 0.99$.

Fig. 3: AFM images of Ga droplet and GaN QD ensembles grown on Si(001): (a) Ga droplets deposited at 550 °C, (b) GaN QDs nitrided at 550 °C, (c) GaN QDs nitrided at 650 °C, and (d)

GaN QDs nitrided at 720 °C. The color-scale ranges displayed are (a) 32.0 nm, (b) 17.7 nm, (c) 18.0 nm, and (d) 33.8 nm. The corresponding size distributions from images (a)-(d) are shown in (e) and the frequency is the percentage of QDs with diameters within a specified range. For Ga droplets, the $d_m$ value is 47±8 nm and the density is 3.7×10$^9$ cm$^{-2}$. For GaN QDs nitrided at 550, 650, and 720 °C the $d_m$ values (densities) are 31±4 nm (5.0×10$^9$ cm$^{-2}$), 26±5 nm (1.2×10$^{10}$ cm$^{-2}$), and multimodal with 18±8 nm and 33±8 nm (1.7×10$^{10}$ cm$^{-2}$) respectively. (a)-(d) are fitted using a log-normal distribution where $R^2 > 0.99$.

Fig. 4: AFM images of Ga droplet and GaN QD arrays grown on Si(111): (a) Ga droplets deposited at 550 °C, (b) GaN QDs nitrided at 550 °C, (c) GaN QDs nitrided at 650°C, and (d) GaN QDs nitrided at 720 °C with a close-up (10× magnification) of a QD cluster shown as an inset. The color-scale ranges displayed are: (a) 29.6 nm, (b) 14.3 nm, (c) 20.2 nm, and (d) 21.7 nm. The corresponding size distributions from images (a)-(d) are shown in (e) and the frequency is the percentage of QDs with diameters within a specified range. For Ga droplets, the $d_m$ value is 47±4 nm and the density is 5.3×10$^9$ cm$^{-2}$. For GaN QDs nitrided at 550, 650, and 720 °C, the $d_m$ values (densities) are 18±4nm (4.2×10$^{10}$ cm$^{-2}$), multimodal with 10±7 nm and 24±7 nm (4.9×10$^{10}$ cm$^{-2}$), and 7±3 nm (1.0×10$^{11}$ cm$^{-2}$) respectively. (a)-(d) are fitted using a log-normal distribution where $R^2 > 0.99$.

Fig. 5: Illustrations of QD nucleation and growth mechanisms on silica and silicon surfaces. In both cases, four steps are shown: (a), (b): "initial" nitridation; (c), (d): Ga depositions; (e), (f), (g): "final" nitridation. On the silica surface, (a) the initial nitridation induces nanoscale surface roughening which subsequently enables Ga droplet formation during (c) the Ga deposition step. During the "final" nitridation step, nitrogen impinges upon the droplets, resulting in QD nucleation. When (f) the final nitridation substrate temperature is at lower temperature (550°C and 650°C), the Ga diffusion length, $\lambda_D$, is much smaller than Ga droplet separation distance, and Ga out-diffusion leads to QD nucleation in between Ga droplets. When the final nitridation substrate temperature is increased to 720°C, the Ga diffusion length, $\lambda_D$, is (e) comparable to the Ga droplet separation, and Ga out-diffusion leads to Ga droplet coarsening. On the silicon surfaces, during (b) the initial nitridation step, impinging nitrogen atoms form patches of surface $Si_xN_y$ layers, which act as droplet nucleation sites during (d) the Ga deposition step. During (g) the "final" nitridation step, nitrogen impinging upon the droplets and the regions of bare silicon surface, enabling QD nucleation at both Ga droplets and $Si_xN_y$ patches. When the final nitridation substrate temperature is increased, the Ga diffusion length, $\lambda_D$, is comparable to the nucleation site separation, and a bimodal distribution of QD sizes is apparent.

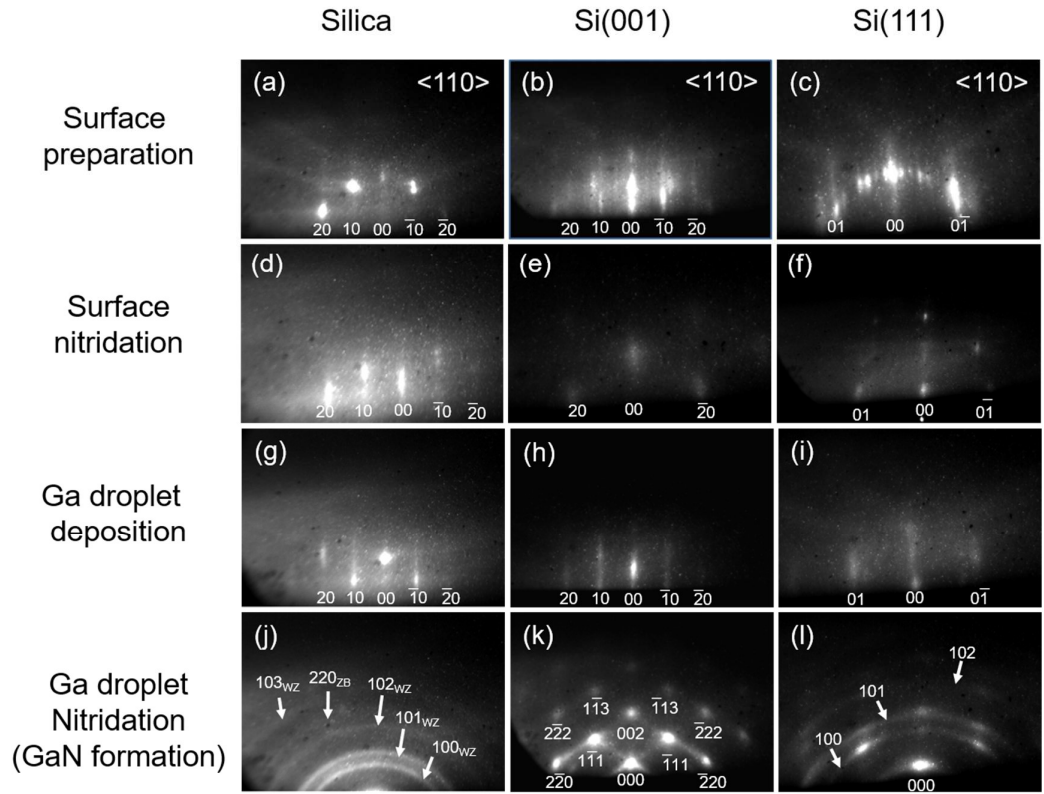

|  | Silica | Si(001) | Si(111) |
|---|---|---|---|
| Surface preparation | (a) ⟨110⟩ | (b) ⟨110⟩ | (c) ⟨110⟩ |
| Surface nitridation | (d) | (e) | (f) |
| Ga droplet deposition | (g) | (h) | (i) |
| Ga droplet Nitridation (GaN formation) | (j) | (k) | (l) |

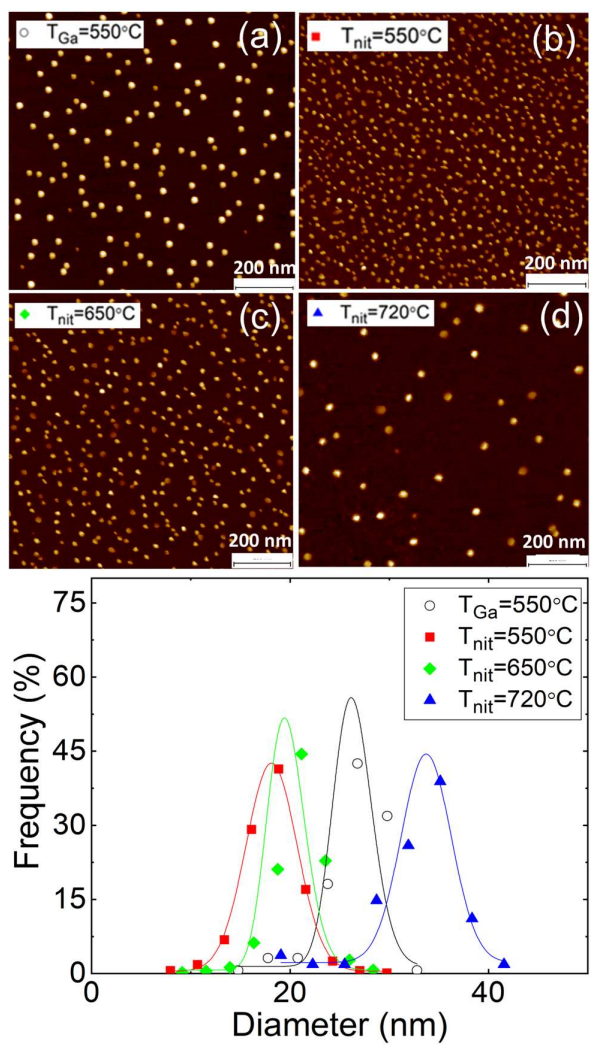

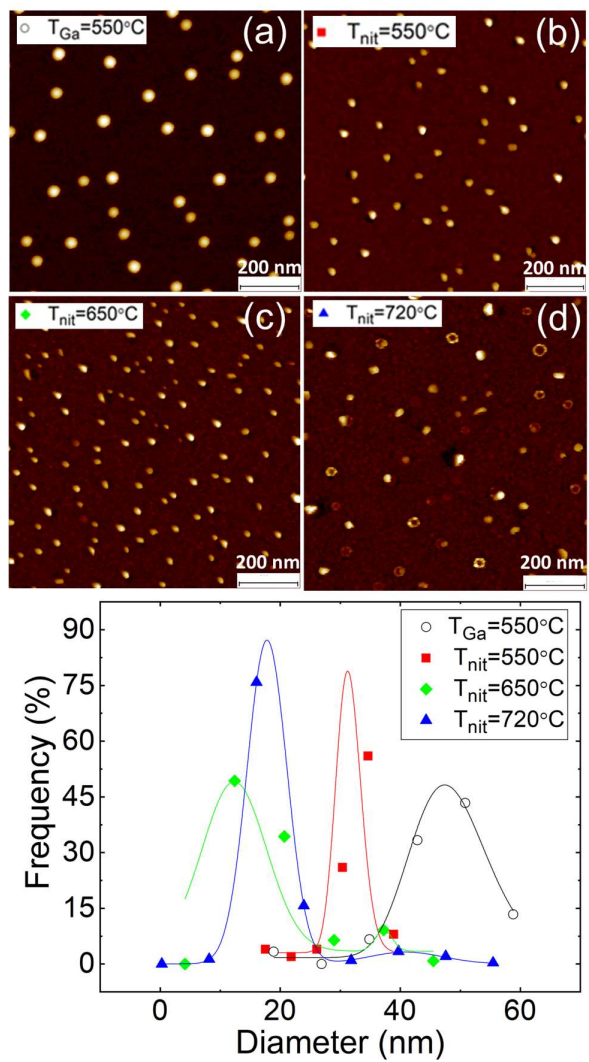

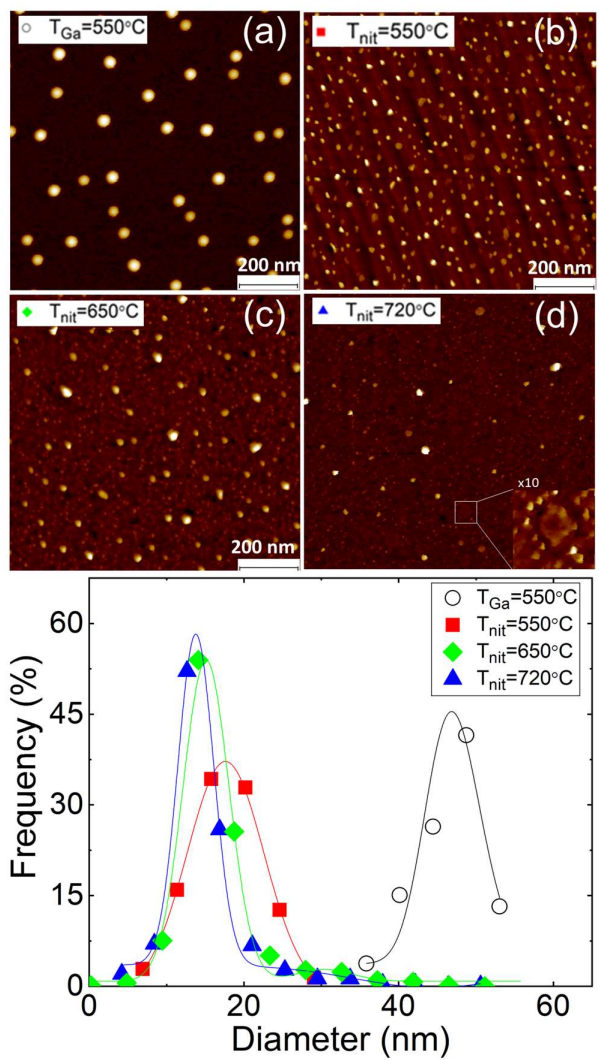

FIG. 5

**Silica**

"initial" nitridation

Ga deposition (a)

"final" nitridation

(b)

Low $\lambda_{Ga}$          High $\lambda_{Ga}$

(c)          (d)

**Silicon**

"initial" nitridation

Ga deposition (e)

"final" nitridation

(f)

(g)

- Gallium
- Nitrogen
- Oxygen
- Surface Silicon

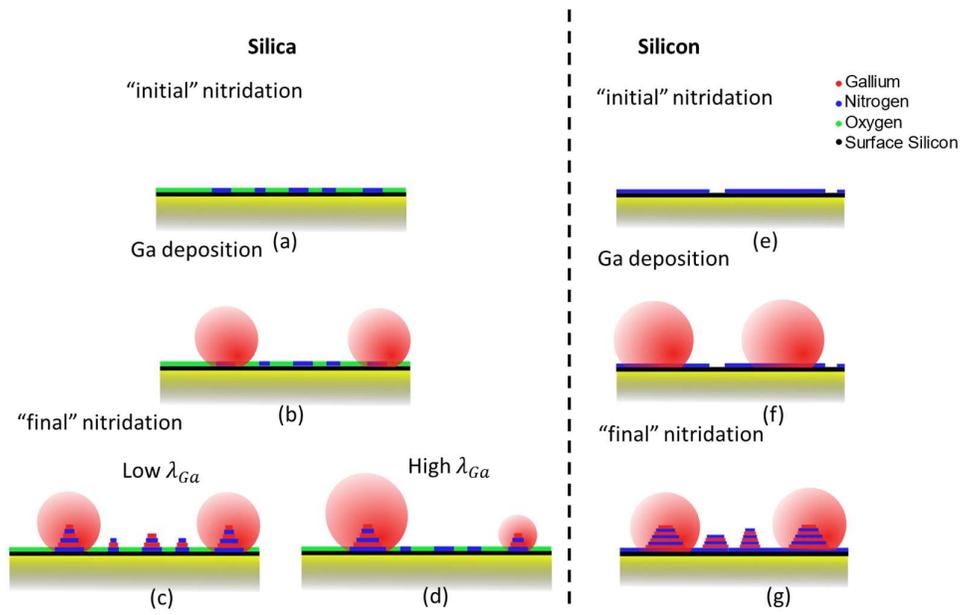

**Mechanisms of GaN quantum dot formation during nitridation of Ga droplets**

H. Lu[1], C. Reese[1], S. Jeon[1], A. Sundar[1], Y. Fan[2], E. Rizzi[1], Y. Zhuo[2], L. Qi[1] and R. S. Goldman[1,3]*

*[1]Department of Materials Science and Engineering*
*University of Michigan, Ann Arbor, MI 48109-2136*
*[2]Key Laboratory for Thermal Science and Power Engineering of the Ministry of Education, Dept.*
*of Energy and Power Engineering, Tsinghua University., Beijing 100084*
*[3]Department of Physics*
*University of Michigan, Ann Arbor, MI 48109-2136*

*rsgold@umich.edu

We describe density functional theory (DFT) calculations of diffusion barriers and adsorption energetics on silicon surfaces. We first discuss calculations of the diffusion barriers of Ga and N adatoms on $SiO_2(001)$, $Si(001)$ and $Si(111)$ surfaces. Using the Nudged-Elastic Band method, reaction paths are constructed for various Ga and N adatom surface diffusion processes. We then discuss calculations of the adsorption energetics of Ga adatoms on nitride/oxidized Si(001) with varying surface coverages.

Ga and N Surface Diffusion Barriers

To compute the Ga and N surface diffusion barriers, we used the Materials Studio CASTEP module,[1,2] with exchange and correlation interactions described using the Generalized Gradient Approximation (GGA) and the Perdew-Burke-Ernzerhof (PBE)[3] functional, and electron-ion core interactions described using ultra-soft pseudopotentials.[4] Geometric optimization was performed using the Broyden-Fletcher-Goldfarb-Shanno (BFGS) optimization algorithm,[5] with convergence criteria including: (a) self-consistent field (SCF) of $5.0 \times 10^{-7}$ eV/atom; (b) energy of $5 \times 10^{-6}$ eV/atom; (c) displacement of $5 \times 10^{-4}$ Å; (d) force of 0.01 eV/Å; and (e) stress of 0.02 GPa. Transition states were determined using the complete linear synchronous transitions (LST)/quadratic synchronous transitions (QST) method[6] with convergence criteria set to 0.05 eV/Å, and confirmed by the Nudged-Elastic Band (NEB) method,[7] with convergence criteria including: (a) energy of $1.0 \times 10^{-5}$ eV/atom; (b) maximum force of 0.05 eV/ Å; and (c) maximum displacement of 0.004 Å.

The energies of crystalline Si and $SiO_2$ cells were converged with $6 \times 6 \times 6$ k points in a Monhorst-pack grid,[8] while the slab models and related adsorption structures were converged with $2 \times 2 \times 1$ k points. The electronic wave functions were expanded in a plane wave basis with a 300 eV cutoff energy. With relaxed surface layer of atoms and fixed buried layers,[9] the vacuum region between slabs was set to 12 Å to avoid interactions among periodic image charges.[10] As shown in Table S1, the calculated lattice parameters are consistent with prior experimental and computational studies,[11,12] confirming the reliability of our calculations. In addition, the computed coordination numbers for Si atoms on Si (001), Si (111) and $SiO_2$ (001) surfaces are 2, 3, and 2, respectively, consistent with values computed in prior first-principles calculations.[15-17]

**TABLE S1:** Table of DFT computed lattice parameters for Si and SiO₂, in comparison with computed[11,12] and experimental[13,14] literature reports.

| Lattice parameters | Prior simulation[11,12] | Experiment[13,14] | This work |
|---|---|---|---|
| $a_{Si}$ (Å) | 5.424 | 5.431 | 5.465 |
| $a_{SiO2(quartz)}$ (Å) | 5.020 | 4.9965 | 5.107 |
| $c_{SiO2(quartz)}$ (Å) | 5.560 | 5.4570 | 5.578 |

As shown in Figs. S1, S1, and S3, the computed diffusion barriers for a Ga adatom on SiO₂(001), Si(001) and Si(111) are 0.32 eV, 0.38 eV and 0.41 eV, respectively.

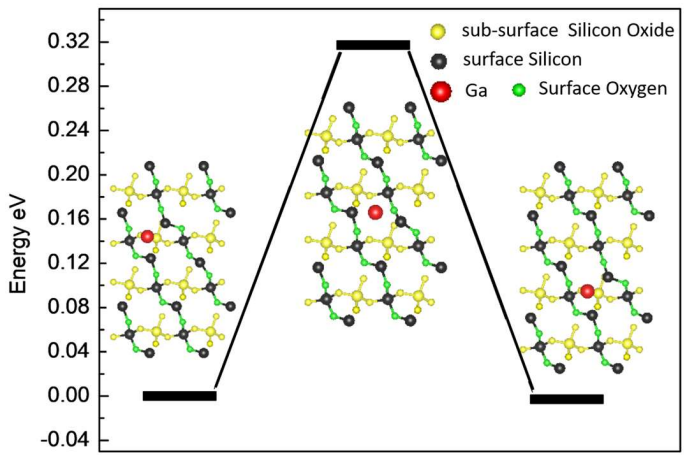

**Fig. S1:** Reaction path for Ga adatom (red) diffusion on SiO₂(001) surface viewed from the top, overlaid on a plot of the energies of the initial, intermediate, and final states. The initial energy state is defined as zero and the surface oxygen, surface silicon, and sub-surface SiO₂ are shown in green, black, and yellow, respectively.

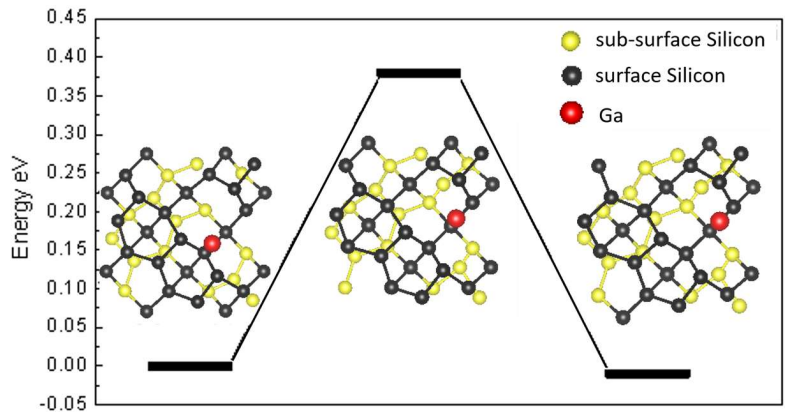

**Fig. S2:** Reaction path for Ga adatom (red) diffusion on Si (001) surface viewed from the top, overlaid on a plot of the energies of the initial, intermediate, and final states. The initial energy state is defined as zero and the surface oxygen, surface silicon, and sub-surface SiO₂ are shown in green, black, and yellow, respectively.

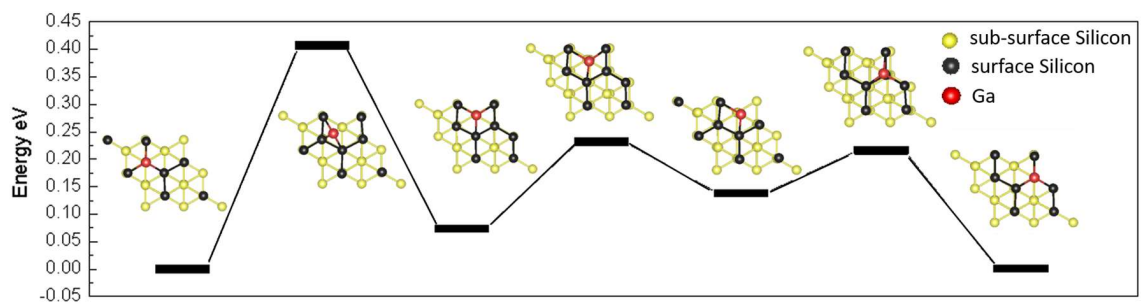

**Fig. S3:** Reaction path for Ga adatom (red) diffusion on Si (111) surface viewed from the top, overlaid on a plot of the energies of the initial, various intermediate, and final states. The initial energy state is defined as zero and the surface oxygen, surface silicon, and sub-surface Si are shown in green, black, and yellow, respectively.

Since the epitaxy process involves the use of a N-plasma source,[18] we assume the predominance of atomic N, as opposed to molecular $N_2$ on the surface. Consequently, as shown in Figs. S4, S5, and S6, the computed diffusion barrier for N adatoms on $SiO_2(001)$, Si(001) and Si(111) are 4.30 eV, 2.64 eV and 3.44 eV, respectively.

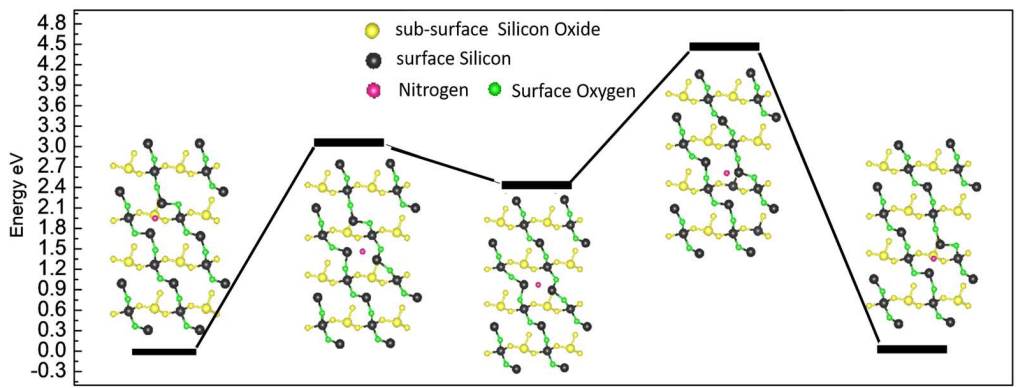

**Fig. S4:** Reaction path for N adatom (pink) diffusion on $SiO_2(001)$ surface viewed from the top, overlaid on a plot of the energies of the initial, several intermediate, and final states. The initial energy state is defined as zero and the surface oxygen, surface silicon, and sub-surface $SiO_2$ are shown in green, black, and yellow, respectively.

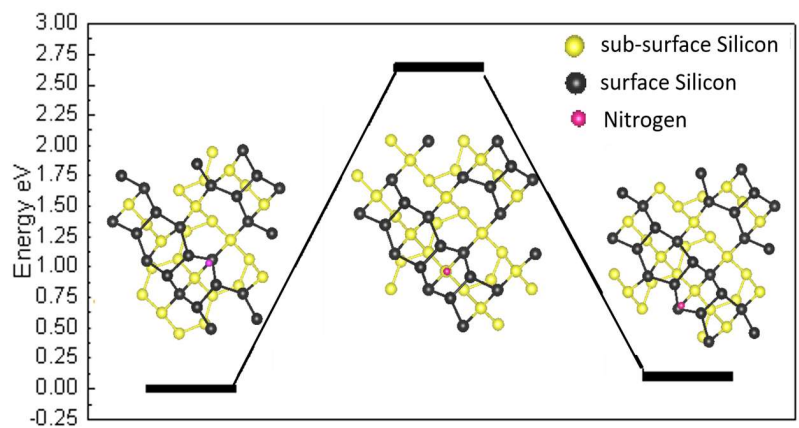

**Fig. S5:** Reaction path for N adatom (pink) diffusion on Si (001) surface viewed from the top, overlaid on a plot of the energies of the initial, intermediate, and final states. The initial energy state is defined as zero and the surface oxygen, surface silicon, and sub-surface Si are shown in green, black, and yellow, respectively.

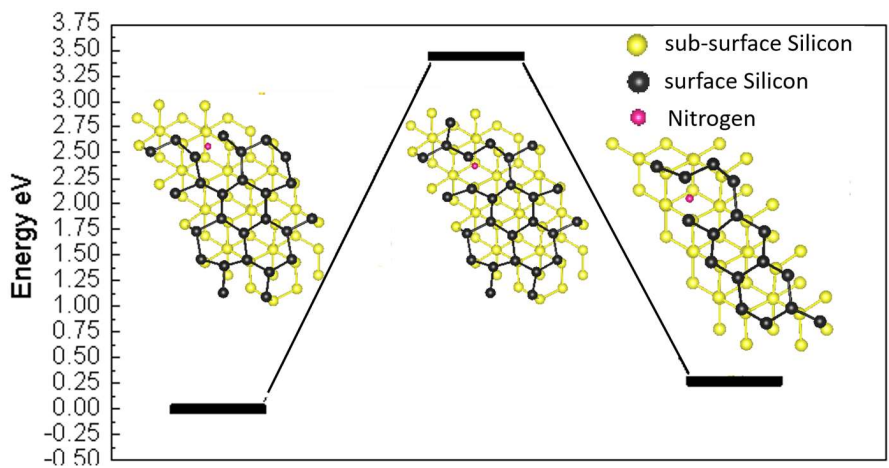

**Fig. S6:** Reaction path for N adatom (pink) diffusion on Si (111) surface viewed from the top, overlaid on a plot of the energies of the initial, intermediate, and final states. The initial energy state is defined as zero and the surface oxygen, surface silicon, and sub-surface Si are shown in green, black, and yellow, respectively.

Since $k_BT \sim 0.0259$ eV when T = 300 K and the pre-exponential factor is on the order of $10^{13}$ s$^{-1}$, the jumping frequency of a Ga adatom is expected to be $\sim 10^6$ s$^{-1}$, covering a region with $\sim 100 \times 100$ nm$^2$ during a 1 second random walk. On the other hand, the jumping frequency of a N atom on is expected to be $\sim 10^{-2}$ s$^{-1}$ even when the temperature reaches 1000 K. Thus, it is expected that Ga adatoms will diffuse along the surfaces, but that N adatoms would be relatively immobile even at the highest growth temperature of 720 °C.

## Ga Adsorption on Nitrided/Oxidized Silicon Surfaces

We also used DFT to compute the adsorption energy of Ga on nitrided and oxidized Si surfaces as follows:

$$E_{ads}^{Ga} = \left[E_{Si} + E_{Si-N(O)-Ga}\right] - \left[E_{Si-N(O)} + E_{Si-Ga}\right] \qquad \textbf{(S1)}$$

where $E_{Si-Ga}$ and $E_{Si-N(O)}$ denote the reference energies for isolated Ga atoms and gas atoms (N or O) on the Si surface; $E_{Si}$ and $E_{Si-N(O)-Ga}$ denote the energies for a clean silicon surface and N(O) bonded to Ga on the silicon surface, respectively. Since first principles methods replicate the thermodynamically-stable (2x1) Si (001) surface reconstruction,[19] but not the (7x7) Si (111) surface reconstruction,[20] our DFT calculations are limited to those for $E_{ads}^{Ga}$ on the Si (001) surface.

All calculations were performed with the Vienna *ab initio* Simulation Package (VASP)[21], using the Perdew-Burke-Ernzerhof[22] (PBE) pseudopotential implemented with the projected augmented-wave[23-24] (PAW) method. The Si (001) surface was modeled using 48 Si atoms in 6 layers, with 16 Å vacuum thickness in the z-direction perpendicular to (001). In the bottom 3 layers, Si ions were kept fixed, while those in the bottom Si surface were H-saturated. In the top 3 layers, the Si ions were mobile and fully relaxed. Following DFT relaxations, the (2×1) reconstruction of the Si (001) surface unit cell was observed, as shown in Fig. S7.

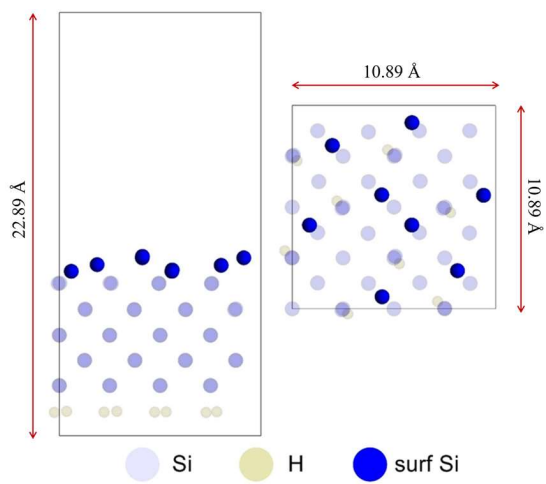

**Fig. S7:** Side and top views of the (2×1) reconstructed Si (001) supercell with 8 surface Si atoms (Dark blue dots labeled as "surf Si"). The bottom surface is saturated by H atoms. By fixing the bottom 3 layers and relaxing atoms in the top 3 layers, the free (001) Si surface undergoes a reconstruction where two near surface Si atoms always come close to each other. The energy of this configuration is $E_{Si}$.

Additional calculations were performed to obtain the Ga adsorption energy as a function of N/O surface coverage on Si(001). For the adsorption of a single N or O atom, $E_{ads}^{Ga}$ is lowest for the bridge site, indicating it is the most favorable thermodynamically. These adsorption geometries of 1 N atom and 1 N + 1 Ga atoms are presented in Fig. S8. Higher surface coverage (2N or 2O atoms) were generated by adsorption of gas atoms at 2 nearby bridge sites (the less stable hollow and lattice sites are not considered). The corresponding adsorption configurations for 2 N atoms and 2 N + 1 Ga atoms are shown in Fig. S9.

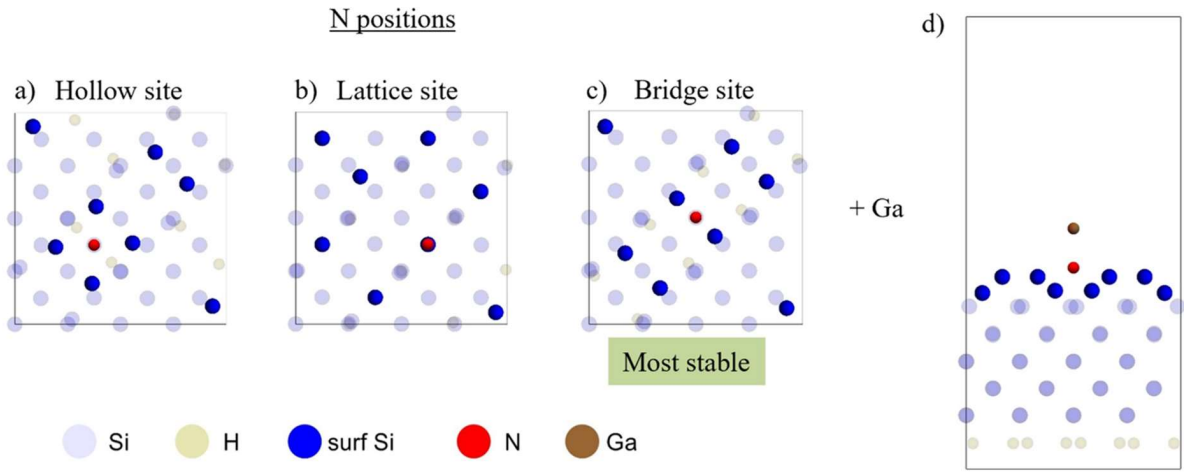

**Fig. S8**: (a) - (c): Top views of adsorption configurations for 1N atom on (2×1) Si (001). The different configurations identified for N are (a) the hollow site with N bonded to 4 surface Si atoms, (b) the lattice site with the N vertically above a surface Si atom, and (c) the bridge site with N bonded to 2 surface Si atoms, respectively. The bridge site is found to be most stable thermodynamically and its energy corresponds to $E_{Si-N}$ in Eq. S1. A Ga atom is added to the surface configuration of (c), resulting a Ga-N bonded configuration as shown in (d) (Side View). The total energy of (d) is accounted for $E_{Si-N-Ga}$ in Eq. S1. Similar structures were obtained for the adsorption of 1 Ga atom on the same Si (001) supercell with 1 O atom.

According to Eq. S1, the calculated adsorption energies of a single Ga atom are +1.25, +1.65, -0.31 and +0.54 eV on (2×1) reconstructed Si (001) surface with 1 N atom, 1 O atom, 2 N atoms and 2 O atoms in the supercell, respectively, as illustrated in Fig. S10. We note that for surface coverage with 1 N or 1 O atom, $E_{ads}^{Ga}$ is positive (i.e. repulsive), indicating a preference for

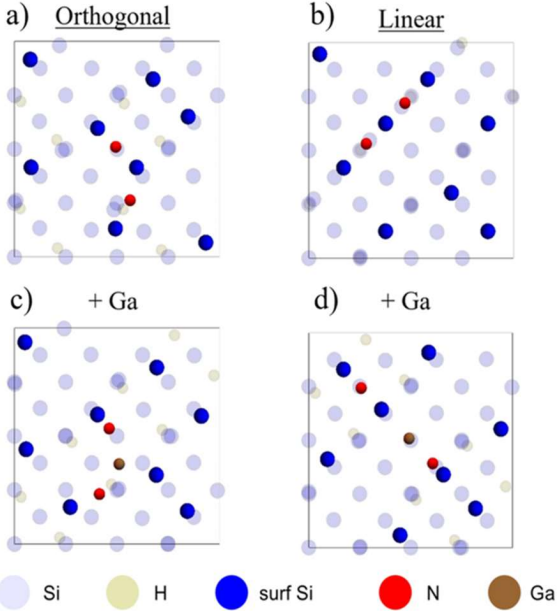

a) Orthogonal
b) Linear
c) + Ga
d) + Ga

Si    H    surf Si    N    Ga

**Fig. S9:** Adsorption energetics of Ga on Si (001) with increasing N surface coverage. 2 N atoms adsorbed in (a) orthogonal or (b) linear bridge sites on the Si (001) surface (hollow and lattice sites are not considered since the bridge site was found to be more stable). In both cases, the central Si atom is bonded to 2 N atoms. The orthogonal arrangement of N atoms in (a) is more stable by 2.0 eV compared with (b). (c) and (d) show adsorption of 1 Ga atom and subsequent bond formation with N. (d) is found to be more stable than (c) by 2.0 eV. Similar structures were obtained for the adsorption of 1 Ga atom on the same Si (001) supercell with 2 O atoms.

isolated Ga adatoms on a clean Si surface without chemical bonding to the N or O atom. For higher N surface coverage, $E_{ads}^{Ga}$ is negative (i.e. attractive), indicating that Ga adatoms prefer to bond to N atoms on the Si surface. However, for higher O surface coverage $E_{ads}^{Ga}$ is positive (i.e. repulsive), indicating it is not energetically favorable for Ga adatoms to bond to O atoms on the oxidized surface. Since $E_{ads}^{Ga}(N) < E_{ads}^{Ga}(O)$, the Ga-N interactions are stronger than Ga-O interactions. Such stronger Ga-N interactions can impede the diffusion on Ga on the surface by pinning Ga atoms to N atoms. Consequently, the lateral Ga diffusivity is expected to be slower on the nitrided than on the oxidized silicon surface. Thus, O-rich regions are expected to provide a pathway for enhanced Ga surface diffusion and nucleation/ripening of GaN quantum dots.

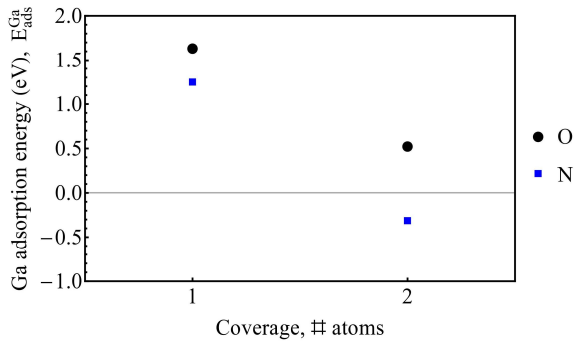

**Fig. S10:** Plot of adsorption energies for a single Ga atom, $E_{ads}^{Ga}$, as a function of the number of adsorbed surface O or N on the (2×1) reconstructed Si (001) surface supercell shown in Fig. S7. Positive (negative) energies are repulsive (attractive). Black dots and blue squares correspond to O and N atoms, respectively.